\documentstyle[prl,aps,epsf,floats,twocolumn]{revtex}
\begin{document}
\draft
\newcommand{\s}{\vskip .2in}
\newcommand{\ev}{\end{verse}}
\newcommand{\hf}{\hfill}
\newcommand{\bv}{\begin{verse}}
\newcommand{\hs}{\hskip .5in}
\newcommand{\be}{\begin{equation}}
\newcommand{\ee}{\end{equation}}
\title
{What angle-resolved\\photoemission experiments
tell about the microscopic theory for
high-temperature superconductors
}
\author{
Elihu Abrahams$^*$ and C.M. Varma$^\dag$}
\address{$^*$Center for Materials Theory, Rutgers University,
136 Frelinghuysen Road, Piscataway, NJ 08854-8019; $^\dag$Bell
Laboratories, Lucent Technologies, 600 Mountain Avenue, Murray
Hill, NJ 07974}
\date{\today}
\maketitle

\begin{abstract}
{\bf ABSTRACT: Recent angular-resolved photemission experiments
on high-temperature
superconductors are consistent with a phenomenological description 
of the normal state of these materials as Marginal Fermi Liquids. The
experiments also provide constraints on microscopic theories.}
\end{abstract}
The discovery of the copper oxide superconducting
materials in 1987 and the intense investigations which
followed have raised some fundamental questions in
condensed matter physics. These superconductors are
characterized by two unexpected features. One is,
of course, their 
unprecedented high transition temperatures
($T_c$). In addition it is
clear that their normal-state properties are not
those of ordinary
metals; they are not consistent with the
traditional Fermi-liquid
quasiparticle picture  which is a cornerstone of
our understanding of the
metallic state.

Many theoretical ideas have been proposed in
response, but it has been
difficult, on the basis of the experimental
evidence, to identify the
correct picture. However, recent angle-resolved
photoemission (ARPES)
experiments show a remarkable consistency with predictions
\cite{mfl}, made in 1989, based on a
phenomenological characterization of these materials as
``Marginal Fermi Liquids" (MFL).
The aim of this communication is to discuss some
aspects of these
experiments and to point out what constraints they
impose on possible
microscopic theories.

The initial motivation for the marginal Fermi liquid
phenomenology was to simultaneously understand two
quite different normal state properties of these quasi-two
dimensional materials: the Raman scattering intensity which
measures the long wavelength response over a
wide range of frequencies and the  nuclear
magnetic relaxation rate of
planar copper nuclei which comes from magnetic
fluctuations of low frequency
and short wavelength. The underlying MFL
assumption was that, at
compositions for which the $T_c$ is highest (``optimal
doping"), a sector of both the charge and
magnetic excitation spectra 
$\chi({\bf q},\omega,T)$ has unusual properties:
i) The sector is momentum ${\bf q}$ independent
over most of the Brillouin zone and
ii) it has a scale-invariant
form, as a function of frequency $\omega$ and 
temperature $T$, so that
${\rm Im}\chi
\propto f(\omega/T)$ as follows:
\begin{eqnarray}
{\rm Im}\chi({\bf q},\omega,T) &= & -N_0 (\omega/T),
\;\;\;\;\;\;  \omega \ll T \nonumber
\\
	& = & -N_0 ({\rm sgn}\omega), \;\;\;\;\,\,
T\ll\omega\ll\omega_c.
\end{eqnarray}
$N_0$ is the density of energy states per unit
volume and $\omega_c$ is a
high-frequency cutoff. A central conclusion
\cite{mfl} is that the single-particle
self energy due to scattering from the excitation
spectrum of Eq.\ (1) has a singular dependence
on frequency and temperature but has 
unimportant momentum dependence. This self energy is
calculated to be
\begin{equation}
\Sigma({\bf k},\omega)=\Sigma_1({\bf k},\omega) +
i\Sigma_2({\bf k},\omega)  =
\lambda[\omega\log\frac{x}{\omega_c} - i\frac{\pi}{2}x],
\end{equation}
where $x \approx \max (|\omega|, T)$ (for
example, $ x= \sqrt{\omega^2+
\pi^2T^2}$) and $\lambda$ is a coupling
constant. The spectra of Eq.\ (1)  could 
actually have
 a smooth dependence
on ${\bf q}$ over a substantial part of the Brillouin zone.
In that case, $\lambda$
acquires a weak dependence on ${\bf k}$. In the experiments
discussed below,  $\lambda$ is in fact constant, within the
experimental error. The most important point is  that
$\Sigma_2$ remains proportional to $x$ all around the Fermi
surface.

In a Fermi liquid, quasiparticles are well-defined
because the single
particle excitation decay rate ($\Sigma_2$) is
small compared to
$\Sigma_1$.  Since in the present case, $\Sigma_2$, the
imaginary part of the self
energy is only logarithmically smaller than $\Sigma_1$, the
real part, at $T=0$, the
appelation  marginal Fermi liquid (MFL) was given \cite{mfl}.
Here, the singular
behavior of $\Sigma$ leads to the result that
there are no Fermi
liquid-like quasiparticles; at $T=0$, their ``residue"
$z = [1 - \partial \Sigma_1/\partial\omega]^{-1}$ 
vanishes at the
Fermi surface $({\bf k=k}_F)$. 

The earliest experimental data revealed that the transport
properties in the normal state of the high-$T_c$
superconductors are unlike those of normal Fermi liquids which
have non-zero $z$. P.W. Anderson suggested \cite{pwa}  that this
experimentally observed anomalous normal-state behavior implies
a non-Fermi liquid with $z=0$.
He developed this idea
based on the separation of charge and spin
energy scales and excitations. The MFL analysis is
different and begins with a
key assumption, Eq. (1), about the particle-hole
excitation spectrum in
{\it both} charge and magnetic (including spin) channels. Eq.\ (2) for
the self energy
follows from Eq.\ (1) with the following consequences:
\begin{itemize}
\item A Fermi surface is defined at $T=0$ as the locus of
${\bf k}$ points where $z\rightarrow 0$ as $(\log \omega)^{-1}$.
\item The single-particle scattering rate $\Sigma_2$
is proportional to $x \approx \max (|\omega|,T)$.
\item The {\it leading order contribution} to $\Sigma_2$
is independent of momentum both perpendicular to and
around the Fermi surface. As discussed above, a weak
smooth momentum
dependence does not alter our conclusions
\end{itemize}

In general, transport (e.g. electrical or
thermal conduction) scattering rates 
have a different frequency and temperature
dependence from that of
the single-particle scattering rate $\Sigma_2$
because the former emphasize backward scattering (large
momentum transfer).
However, the combination of momentum independence of
$\Sigma$ and the linear
dependence on  $x$ of $\Sigma_2$ in the MFL form
of the self energy, Eq.\
(2), leads to a resistivity which is linear in
$T$, as is seen in experiments on optimally-doped
materials. Other
consequences of the MFL phenomenology include a
temperature-independent
contribution (with logarithmic corrections) to the
thermal conductivity; 
an optical conductivity which falls
off as with frequency as $\omega^{-1}$ 
(with logarithmic corrections), more slowly than the
$\omega^{-2}$ dependence of the familiar Drude form; a
Raman scattering intensity
$\propto\max(\omega ,T)/T$; a
$T$-independent contribution to the
copper nuclear spin relaxation rate;
and (in some geometries) a linear in bias voltage
contribution to the single-particle tunneling rate.
The MFL phenomenology has often been used to fit
experiments and it is found that the
behavior of response functions is generally
consistent with MFL as
expressed in Eq.\ (1). 

Nevertheless, although there were some direct
indications of the
correctness of Eq. (2) in  early angular resolved
photoemission (ARPES) measurements
\cite{olson}, the MFL
behavior of the single-particle excitation
spectrum [i.e., Eq.\ (2)] was
not adequately confirmed. ARPES experiments measure
the single-particle
properties directly, in contrast to response
functions which are
governed by joint two-particle (that is,
particle-hole) properties. The
quantity determined in ARPES experiments
is the single-particle
spectral function ${\cal A}({\bf k},\omega)$ which
depends on the self
energy as follows:
\begin{equation}
{\cal A}({\bf k},\omega) =
-\frac{1}{\pi}\frac{\Sigma_2({\bf
k},\omega)}{[\omega - \varepsilon_k -
\Sigma_1({\bf k},\omega)]^2 
+ [\Sigma_2({\bf k}, \omega)]^2}.
\end{equation}
The MFL behavior of the single-particle
excitations has now  been
verified convincingly in the new ARPES experiments
of Valla et al
\cite{valla} at Brookhaven National Laboratory and by
Kaminski et al \cite{kaminski} at Argonne National Laboratory.
In the past, such measurements have
been limited by  energy and momentum resolution
and large experimental
backgrounds in the energy distribution
measurements at fixed momentum
(EDCs). These problems are now being overcome as
new detectors have come
on line. In particular,
Valla et al \cite{valla} have taken advantage of improved
resolution to measure, on
optimally doped
Bi$_2$Sr$_2$CaCu$_2$O$_{8+\delta}$, in addition to
EDCs,
momentum distributions at fixed energy (MDCs). In
this way, the
frequency dependence of the single-particle
spectral function ${\cal A}({\bf k},\omega)$ is measured at
fixed ${\bf k}$ (EDC) and also
the ${\bf k}$
dependence at fixed
$\omega$ (MDC). 

It follows from Eq.\ (3) that if the self-energy
$\Sigma$ is momentum
independent perpendicular to the Fermi surface,
then an MDC  scanned along ${\bf k}_{\rm perp}$
for $\omega \sim 0$
should have a  lorentzian shape plotted against
$({\bf k-k}_F)_{\rm perp}$ 
with a width
proportional to
$\Sigma_2(\omega)$ and the $\Sigma_2$ found in
this way from MDCs should
agree with that found by fitting EDCs.
Furthermore, for  an MFL, the
width should be proportional to
$x = \max(|\omega|,T)$, where
$\omega$ is measured from the chemical potential.
This behavior has now been verified by
Valla et al \cite{valla}.
The fits of the MDC's at $\omega=0$ to a lorentzian 
are shown in Fig.\ 1A. 
Fig.\ 1B shows the linear variation of the 
width of the lorentzian with temperature.
\vskip -.3in
\begin{figure}[htb]
\centering
\hskip -1.4truein
\epsfxsize=4.8truein
\epsffile{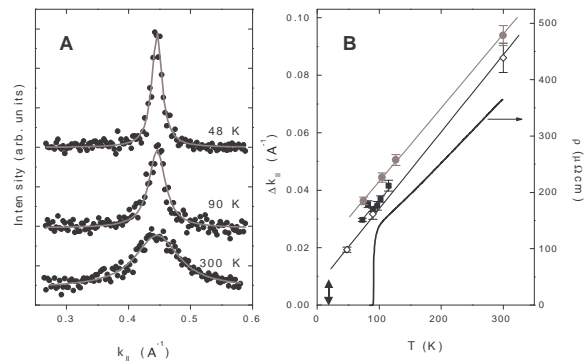}
\vskip -1.5in
\caption{A. Momentum distribution curves for different
temperatures. The solid lines are lorentzian fits. B. Momentum
widths of MDCs for three samples (circles, squares and
diamonds). The thin lines are $T$-linear fits. The resistivity
(solid black line) is also shown. The double-headed arrow shows
the momentum resolution of the experiment. Figures courtesy
of P.D. Johnson (Brookhaven).}
\end{figure}

Preliminary data from both the Brookhaven \cite{m2s} and
Argonne \cite{private} groups also show that the contribution
to $\Sigma_2$  which is proportional to $x$ as determined by
scans perpendicular to the Fermi surface is very weakly
dependent on
${\bf \hat{ k}}_F$; i.e. it varies only weakly with the angle
on the Fermi surface. It is important to notice that there is
no evidence of a $T^2$ contribution to Im$\Sigma$ in the
neighborhood of the Fermi surface anywhere in the Brillouin
zone; the temperature dependent part is always $T$-linear.
Phenomenological ideas
which seek to explain the transport anomalies
in the cuprates
on the basis of hot and cold spots on the
Fermi surface are not consistent with this
experimental finding since they are based on having a $T^2$
behavior in the (1,1) direction and a $T$ behavior in the (1,0)
direction.

The Argonne group has plotted the EDCs together with fits to
the MFL spectral function of Eq.\ (3) at over a dozen ${\bf
k}$-points between the (1,1) and (1,0) directions in the
Brillouin zone. Im$\Sigma(\omega)$ is taken to be of the form
$\Gamma({\bf \hat{k}}) +
\lambda({\bf \hat{k}}_F)\omega$. $\Gamma$ represents an
impurity contribution (see below). We show two typical examples
in Figs.\ 2A and 2B. These display respectively the results at
the Fermi surface in the (1,0) and the (1,1) directions in the
Brillouin zone which give the widest ${\bf \hat{k}}$ variations
of $\Gamma$ and $\lambda$. These self-energy parameters for the
fit are given in the figure caption.
\begin{figure}[h]
\hskip 1in\epsfxsize=3truein 
\epsffile{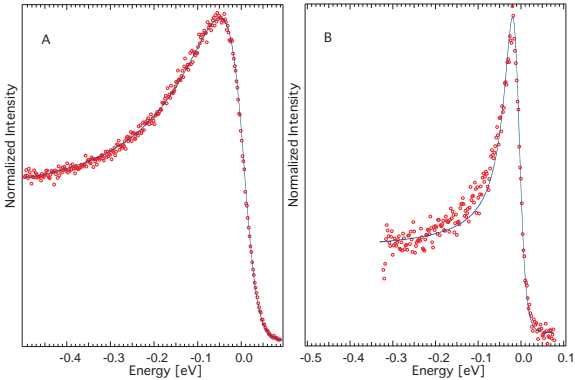}
\vskip .3in
\caption{Fits of the MFL self energy
$\Gamma + \lambda\hbar\omega$ to the experimental data.
Energies are in meV, with estimated uncertainties of $\pm$15\%
in $\Gamma$ and $\pm$25\% in $\lambda$. A., the (1,0)
direction, $\Gamma = .12,\, \lambda=.27$ and B., the (1,1)
direction, $\Gamma = .035,\, \lambda=.35$. Figures courtesy of
A. Kaminski (Argonne). }
\end{figure}

Thus, the results for MDC's as well as EDC's may be summarized by the
following expression:
\begin{equation}
{\rm Im}\Sigma({\bf k},\omega;T) = \Gamma({\bf {\hat
k}}_F)  + {\rm
Im}\Sigma_{MFL}(\omega;T).
\end{equation}
The first term on the right-hand side of Eq.\ (3)
is independent of
frequency and temperature and is properly
considered as the scattering
rate due to static impurities. This can depend on
${\bf \hat{k}}_F$, the
direction of ${\bf k}$ around the Fermi surface,
as explained below. The
second term is the MFL self energy of Eq.\ (2), 
a function only of $x={\rm max}(|\omega|,T)$; however a weak
dependence on ${\bf \hat{k}}$ is possible, as discussed
earlier. There may as well be additive analytic contributions
of the normal Fermi-liquid type.

The dependence
of the impurity scattering on
$\hat{k}_F$ can be understood by the assumption that in
well-prepared cuprates the impurities lie between the  CuO$_2$
planes and therefore give rise to small-angle scattering (small
momentum transfer) only. If the distance
 of the impurities to the CuO plane is $D$, then the
characteristic scattering angle is $\delta\theta\sim
(2k_FD)^{-1}\,\, [\sim {\rm O}(a/d)]$, where $a$ and $d$ are
the in-plane and
$c$-axis lattice constants). The single-particle impurity
scattering rate $\Gamma_{\bf k}$ at a point ${\bf k}$ is then
proportional to
$\delta\theta$ about that point.  For small
$\delta
\theta$, this gives the scattering rate $\Gamma_{\bf k}$  
proportional to the local density of states at that ${\bf k}$.
It is known from both band
structure calculations and ARPES experiments that the local
density of states is about an order of magnitude
larger in the ($\pi,0$) direction than in the
($\pi,\pi$) direction. Therefore, we expect the
impurity contribution to increase as one turns
${\bf {\hat k}}$ around the Brillouin zone toward
the $(\pi,0)$ region. This
is seen in preliminary data from the Brookhaven group
as well as the Argonne group
which also show that the MFL contribution
proportional to $x$ has only weak 
dependence on
${\bf {\hat k}}_F$, if any \cite{m2s,private}.

This assumption about the nature of impurity scattering also 
explains the well-known fact that the impurity contribution
to the resistivity of well-prepared optimally doped
cuprates is anomalously small.
An impurity
contribution to the single-particle scattering rate
does not
necessarily appear in the transport scattering rate since
the latter depends only on large  momentum transfers. 
Thus the transport rate, emphasizing as it does the
large-angle scattering, is proportional to $\delta\theta^3$;
thus a factor $\delta\theta^2$ (perhaps as small as .04) smaller
than
$\Gamma$. In fact, it is known that impurities like Zn 
which replace Cu in the plane
give a large contribution to the resistivity \cite{Zn}.

Some properties at optimum doping, for example the temperature
dependence of the Hall resistivity, do not follow from the MFL
self energy, Eqs.\ (2,4). Although the ARPES experiments
indicate that the MFL form of the self energy is necessary, it
is not sufficient to characterize all the normal state
properties at optimum doping.

The negligible (or weak) experimentally observed momentum
dependence of the singular part of
the single-particle self energy constrains
theories of the normal state
of the cuprate superconductors.  We have discussed
how the MFL
phenomenology is consistent with the experiment.
There are at least three
classes of theories in which (in contrast to MFL) the self energy
has strong momentum
dependence: i).\ Theories which involve strongly
momentum-dependent
couplings such as antiferromagnetic spin
fluctuation exchange
\cite{pines}.
ii).\ Theories based on breaking of
translational symmetry, such as stripe or
charge-density wave scenarios
\cite{emery,cdw}. iii).\ Theories based on an
extension from one to two
dimensions of anomalous non-Fermi liquid behavior
as described
in the Luttinger liquid formulation \cite{pwa2}.
In the
first case, coupling involving an excitation whose
spectrum is
peaked in one part of the Brillouin zone
necessarily leads to a
momentum-dependent self energy. In the second
case, the
breaking of translation invariance also results in
momentum dependence of
all quantities. The third case, the Luttinger
liquid, has non-analytic dependence of the spectral
function on momentum, that is on ${\bf k-k}_F$, at
$\omega=0$ that is the same 
as the $\omega$ dependence at $k=k_F$ \cite{lut}. Thus,
should the new ARPES results 
prove robust, these three
classes of theories would require important
modifications. 

In the theory of the MFL, the
$\max(|\omega|,T)$ dependence of both the
single-particle and transport
rates comes about from a particle-hole
fluctuation spectrum which
is both scale-invariant in frequency and has negligible
momentum dependence as in
Eq.\ (1). A direct experimental verification of the basis for MFL, that is
of a scale-invariant fluctuation spectrum as in Eq.\ (1) is
as yet incomplete. While such a spectrum
is observed at long wavelengths in Raman
scattering, it has not yet been clearly identified at larger
momentum. Experimental methods to measure the charge
fluctuation spectrum at large momentum are not well developed.
The magnetic part of the spectrum (due to both orbital and spin
fluctuations) is in principle observable in neutron scattering.
However, the signal to noise and background problems for a
featureless inelastic spectrum spread over 
$\omega_c \sim 0.5$eV are formidable.
At the same time, the fact that a magnetic fluctuation
spectrum of the MFL form accounts for the
longitudinal NMR relaxation rate of planar copper nuclei near
optimum doping does give support to Eq.\ (1) for large momenta.

It is not clear to us how the observed ARPES spectra (and the
fact that the transport scattering rate has the same
temperature  and frequency dependence as the single particle
scattering rate) can come about except in a theory which
results in the scale-invariant form essentially as specified by
Eq.\ (1). It is suggestive that a scale-invariant fluctuation 
spectrum generally arises in a region about a quantum critical
point (QCP).  The momentum-independence of Eq.\ (1) specifies
that if a QCP underlies the MFL behavior, then the associated
dynamical critical exponent 
$z_d \rightarrow \infty$. As stressed elsewhere \cite{varma},
the experimental phase diagram in the temperature-doping plane
is consistent with the existence of a QCP at a doping near that
of the highest critical temperature. A microscopic theory for a
QCP with properties similar to the MFL spectrum of Eq. (1) has
been presented and shown  also to promote $d$-wave
superconductivity \cite{varma}. Some unique predictions of the
theory await experimental tests.

This work was supported in part by
National Science Foundation Grants DMR9632294 and DMR9976665
to E.A.

\end{document}